\documentclass[twocolumn,prl,superscriptaddress]{revtex4-1}

\usepackage[english]{babel}
\usepackage{boldline,multirow,xcolor,colortbl}
\usepackage[ansinew]{inputenc}
\usepackage{times}
\usepackage{graphicx}
\usepackage{graphics}
\usepackage{amsmath}
\usepackage{amsfonts}
\usepackage{amssymb}
\usepackage{dsfont}
\usepackage{epstopdf}
\usepackage{makeidx}
\usepackage{subfigure}
\usepackage{color}
\usepackage{pgf}
\usepackage{bm}
\usepackage{tikz} 
\usepackage[normalem]{ulem}

\addto\captionsenglish{}

\begin{document}

\title{Multiband Effects in Equations of Motion of Observables Beyond the Semiclassical Approach}

\author{Troy Stedman}
\affiliation{Department of Physics, University of South Florida, Tampa, Florida 33620, USA}

\author{Carsten Timm}
\affiliation{Institute of Theoretical Physics and Center for Transport 
and Devices of Emergent Materials, Technische Universit{\"a}t Dresden, 
01062 Dresden, Germany}

\author{Lilia M. Woods}
\affiliation{Department of Physics, University of South Florida, Tampa, Florida 33620, USA}
\date{\today}

\begin{abstract}

The equations of motion for the position and gauge invariant crystal momentum are considered for multiband wave packets of Bloch electrons. 
For a localized packet in a subset of bands well-separated from the rest of the band structure of the crystal, one can construct an effective electromagnetic Hamiltonian with respect to the center of the packet. 
We show that the equations of motion can be obtained via a projection operator procedure, which is derived from the adiabatic approximation within perturbation theory. 
These relations explicitly contain information from each band captured in the expansion coefficients and energy band structure of the Bloch states as well as non-Abelian features originating from interband Berry phase properties. 
This general and transparent Hamiltonian-based approach is applied to a wave packet spread over a single band, a set of degenerate bands, and two linear crossing bands. 
The generalized equations of motion hold promise for novel effects in transport currents and Hall effect phenomena.

\end{abstract}

\maketitle


{\it Introduction - } Semiclassical theory has been a useful tool in describing various transport phenomena in materials and composites. 
This powerful theory assumes that under external fields, electrons essentially behave as free particles provided the energy band structure is used as the dispersion. 
Recent developments have shown that this picture is not adequate and a more rigorous treatment is needed. 
One of the most striking examples is the correction to the usual quasiparticle group velocity determined by the energy band structure \cite{Xiao-2010}. 
Such an anomalous velocity correction comes from the Bloch states Berry curvature, which can lead to important modifications in many phenomena. 
A number of new features, such as the internal anomalous and spin Hall effects, anomalous thermoelectricity, and intrinsic magnetic moments of electronic wave packets among others  have been demonstrated as a result of these new developments \cite{RevModPhys-2010, RevModPhys-2015, Culcer, Niu-2008, Luttinger, Niu-2002,Fiete,Adrian}. 
This transport formalism has also been applied to light propagation  in photonic materials, where the Berry phase effects have been shown experimentally \cite{NatPhotonics}. 

The anomalous velocity from the Berry phase is important in semiclassical transport and has been considered in several studies for the time dynamics in terms of characteristic equations of motion (EOM) of Bloch electrons in an electromagnetic field \cite{Niu.95,Niu.96,Niu.99,Shindou-2005,Culcer,Gosselin,Xiao-2010}. 
Several phenomena linked to this anomalous velocity are especially pronounced in materials with nontrivial topology, such as Weyl and Dirac materials, where an array of intrinsic Hall effects have been demonstrated \cite{Xiao-2010,RevModPhys-2018,Wehling}. 
The EOM description typically assumes that the wave packet spreads out over several unit cells of a lattice and the electromagnetic waves are taken in the long-wavelength approximation with a much larger spread than the packet. 
For such a situation, one can construct an effective Hamiltonian as a perturbative series with respect to the location of the center of the wave packet in real space \cite{Niu.95,Niu.96,Niu.99,Shindou-2005}. 
 A key issue here is the number of bands comprising the wave packet in reciprocal space. 
Several works have considered wave packets extending over a single band \cite{Niu-2002,Niu-2008,Niu.95,Niu.96,Niu.99}, which is in line with the assumption from semiclassical theory that a single band far away from band degeneracies and crossings in the band structure dominates the contributions to transport. 
A limited number of studies have examined the dynamics of wave packets that extend over multiple degenerate bands. 
For example, such multiband situations have been found to influence various types of topological current transport, like parity polarization currents \cite{Shindou-2005}, or the time dynamics of the spin degree of freedom for degenerate bands \cite{Culcer}. 

The general scenario of describing the EOM for a wave packet extending over more than one band needs a thorough investigation, however. 
This is especially relevant for topological materials with spin textures, where interband effects can lead to non-Abelian gauge fields arising from an SU(N) invariance of the dynamics in a subspace of N degenerate bands. 
In this context, several other important issues need to be studied. 
For example, in deriving the wave packet dynamics, one typically invokes a projection procedure of operators onto a subset of the wave packet bands. 
This could be onto a single band as is the case in \cite{Niu.99,Gosselin} or a subset of degenerate bands as is the case in \cite{Culcer,Shindou-2005}. 
A thorough justification of such a projection is for the most part lacking, however. 
Another issue is that the wave packet dynamics derived from the effective Hamiltonian is a perturbative series. 
The notion of order coming from this perturbative series and its effect on the EOM of the position and gauge invariant crystal momentum needs further understanding. 
EOM  derived for bands of different dispersions and presented in a straightforward manner  is also unavailable. 
Thus more work is needed in order to understand multiband effects in the EOM.

Here we present a direct approach for deriving EOM for the position and gauge invariant crystal momentum of multiband Bloch electron wave packets in the long-wavelength limit. 
Within this method, we use a perturbative series of the general electromagnetic Hamiltonian with respect to the location of the wave packet \cite{Xiao-2010,Niu.99}. 
It is shown that for slowly time varying fields and a multiband subset of bands that are degenerate or nearly degenerate at some point in reciprocal space near the Fermi level and sufficiently separated in energy from the rest of band structure, one can apply perturbation theory with adiabaticity. From this the projected operators onto this band subset follow naturally. 
Full expressions for the EOM in terms of multiband quantities, like Berry curvatures and connections, are given explicitly and applied to the cases of N degenerate bands and a two-band model with a linear dispersion. 
In addition to its transparency, this method gives the basic framework to study multiband electron wave packet dynamics, which is to be used in future quantum kinetic transport models to capture multiband effects in transport properties. 

{\it Theoretical Model - } In the presence of electromagnetic fields, quantum mechanical processes can be described using the Hamiltonian $H(\mathbf{p}-e \mathbf{A}(\mathbf{x}),\mathbf{x})=\frac{1}{2m}(\mathbf{p}-e \mathbf{A}(\mathbf{x}))^2+V(\mathbf{x}) +e\varphi(\mathbf{x})$, where $\mathbf{p}$ is the momentum operator and $\mathbf{A}(\mathbf{x})$ and $\varphi (\mathbf{x})$ are the vector and scalar potentials, respectively. 
In a solid, the fields propagate in a periodic environment set by the crystal lattice with Bloch states giving the natural basis in reciprocal space. 
Transport in solids, as captured in the Boltzmann equation approach \cite{Ashcroft}, involves semiclassical particles represented as wave packets \cite{Shindou-2005}
\begin{equation} \label{eq 2.1}
\psi = \sum_{n} \int d^3\mathbf{q} c_n(\mathbf {q})\psi_n(\mathbf {q}), c_n(\mathbf{q}) = \sqrt{\rho (\mathbf{q} - \mathbf{q}_c) } z_n ( \mathbf{q} ),
\end{equation}
where $\psi_n(\mathbf{q}) = e^{i\mathbf{q}\cdot\mathbf{x}}u_n(\mathbf{q})$ are Bloch states with Bloch functions $u_n(\mathbf{q})$ for wave vector $\mathbf{q}$ in the $n^{th}$ band with expansion coefficients $c_n$ with $ z_n = a_n e^{-i\gamma_n (\mathbf{q})}$. 
Thus the wave packet is characterized by a phase $\gamma_n(\mathbf{q})$ and an amplitude $\rho (\mathbf{q} - \mathbf{q}_c)$, which specifies the localization of the packet about the expectation value of its center $\langle \mathbf{q} \rangle=\mathbf{q}_c$ in reciprocal space. 
The probability that the particle is found in the $n^{th}$ band is $a_n^2$.

We assume the packet is well-localized in reciprocal space and the time scale of changing its shape is larger than the time scale of the dynamics of its center. In Fig. 1, we schematically show how such a packet spreads out over several unit cells in real space due to its narrow localization in reciprocal space.
For transport, one typically works with single band wave packets in the long-wavelength approximation, where the wavelengths $\lambda$ of the fields Fourier components are much longer than the spread in real space $\Delta x$, thus $\epsilon=\frac{\Delta x}{\lambda} \ll 1$.  
Neglecting the wave packet spread in reciprocal space, one then obtains the well-known semiclassical EOM {\cite{Ashcroft,Blount,Zak,Littlejohn}
\begin{equation} \label{eq 2.2}
\dot{ \langle \mathbf {x} \rangle}=\frac{1}{\hbar}\nabla\mathcal{E}(\mathbf {k}_c), \dot{ \langle \mathbf {k} \rangle} = \frac{e}{\hbar}(\mathbf E+ \dot{ \langle \mathbf {x} \rangle} \times \mathbf B), 
\end{equation}
where $ \langle \mathbf {x} \rangle = \mathbf{x}_c$ and $ \langle \mathbf {k} \rangle = \mathbf{k}_c$ are the expectation values of position and gauge invariant crystal momentum $\mathbf{k} = \mathbf{q} - \frac{e}{\hbar}\mathbf{A}$ respectively. 
Here $\nabla\mathcal{E}(\mathbf {k}_c)$ is the reciprocal space gradient of the single band dispersion of the bare crystal Hamiltonian $H_0 (\mathbf{p},\mathbf{x}) = \frac{\mathbf{p}^2}{2m} + V(\mathbf{x})$ with $H_0 (\mathbf{p}+\hbar\mathbf{q},\mathbf{x}) u_n (\mathbf{q}) = \mathcal{E}_n(\mathbf{q}) u_n (\mathbf{q})$
($\nabla$ is understood as differentiation with respect to $\mathbf {k}_c$) and $\mathbf{E}$ and $\mathbf{B}$ are the external electric and magnetic fields respectively. 
The EOM can then be utilized within the Boltzmann equation to determine electric and heat currents in a given material \cite{Ashcroft}.

\begin{figure}
\includegraphics[width=0.22\textwidth]{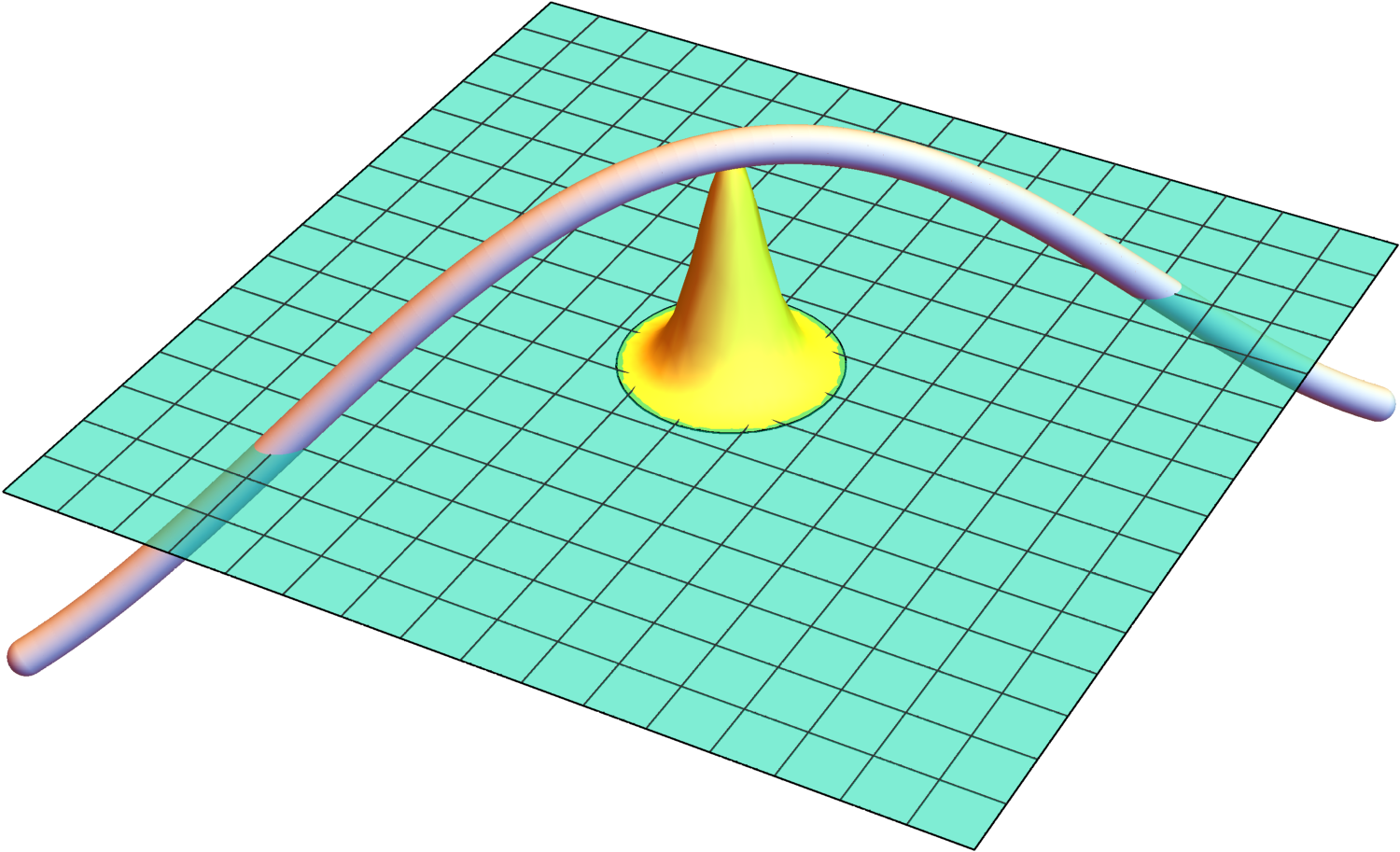}
\caption{Schematics of a wave packet (yellow) spread over several unit cells in a solid, depicted as squares in the plane. 
The long wave-length nature of the electromagnetic wave (pink curve) is also depicted, showing that it is much larger than the packet spread.}
\label{fig1}
\end{figure}

It has been recognized, however, that the interplay between the time dynamics of the fields and the material band structure has to be re-examined within the semiclassical approximation \cite{Xiao-2010,Niu.99,Luttinger,Niu.95,Niu.96}. 
Specifically, due to the long-wavelength approximation, one can effectively use a power series of the field potentials in the Hamiltonian about the real space wave packet center $\mathbf{x}_c$  \cite{Niu.99}. 
Retaining only up to the first order in this series, one finds 
\begin{equation} \label{eq 2.3}
H \approx H_c(\mathbf{p} - e\mathbf{A}_c , \mathbf{x} ; \mathbf{x}_c)+\frac{1}{2}\frac{\partial H_c}{\partial \mathbf{x}_c} \cdot (\mathbf{x} - \mathbf{x}_c) +h.c.
\end{equation}
Here $H_c(\mathbf{p} - e\mathbf{A}_c , \mathbf{x} ; \mathbf{x}_c) = H_0(\mathbf{p} - e\mathbf{A}_c , \mathbf{x}) + e\varphi_c =\frac{1}{2m}\left(\mathbf{p}-e\mathbf{A}_c\right)^2+V(\mathbf{x})+e\varphi_c$, $\mathbf{A}_c = \mathbf{A} (\mathbf{x}_c)$ is the vector potential and $\varphi_c = \varphi (\mathbf{x}_c)$ is the scalar potential, which are all functions of the center of the wave packet $\mathbf{x}_c$. 
The last two terms in (\ref{eq 2.3}) determine the first order perturbation of the linearized Hamiltonian in the long-wavelength limit via the parameter $\epsilon=\frac{\Delta x}{\lambda}$. 
This is realized by noting that $\frac{\partial H_c}{\partial \mathbf{x}_c} = \sum_j \frac{\partial H_c}{\partial A_{c,j}} \frac{\partial A_{c,j}}{\partial \mathbf{x}_c}$ results in a $1/\lambda$ factor from the Fourier transformed vector potential (see (S1) in the Supplementary Information), while $(\mathbf{x} - \mathbf{x}_c)$ determines the wave packet spread in real space $\Delta x$. 
Using a Lagrangian formalism within the semiclassical single band approximation \cite{Niu.95,Niu.96,Niu.99}, the EOM are found as
\begin{equation} \label{eq 2.4}
\dot{ \langle \mathbf {x} \rangle}=\frac{1}{\hbar}\nabla\mathcal{E}_M(\mathbf {k}_c) - \dot{ \langle \mathbf {k} \rangle} 
\times \bar {\mathbf{\Omega}}_n, 
\dot{ \langle \mathbf {k} \rangle} = \frac{e}{\hbar}(\mathbf E+ \dot{ \langle \mathbf {x} \rangle} \times \mathbf B),
\end{equation}
where $\bar {\mathbf{\Omega}}_n = \nabla \times \bar {\mathbf{\mathcal{A}}}_n$ is the Berry curvature and $ \bar{\mathbf{\mathcal{A}}}_n = i \langle u_n (\mathbf{k}_c) |  \nabla | u_n (\mathbf{k}_c) \rangle$ is the Berry connection with Bloch functions $u_n (\mathbf{k}_c)$ for the $n^{th}$ band. 
The energy $\mathcal{E}_M(\mathbf {k}_c) = \mathcal{E}(\mathbf {k}_c) - \bar {\mathbf{M}}_n \cdot \mathbf{B} $ is the lattice band dispersion with a contribution from a reciprocal space magnetic moment $\bar {\mathbf{M}}_n = - \frac{ie}{2\hbar} \langle  \nabla u_n(\mathbf{k}_c) | 
\left( \mathcal{E}(\mathbf {k}_c)  - H_0 (\mathbf{p}+\hbar\mathbf{k}_c,\mathbf{x}) \right) \times |  \nabla u_n(\mathbf{k}_c) \rangle $ of the single band  packet
that couples to the magnetic field $\mathbf{B}$.

One notes that  $H_c(\mathbf{p} - e\mathbf{A}_c , \mathbf{x} ; \mathbf{x}_c)$ in the Hamiltonian from (\ref{eq 2.3}) leads to the standard well-known EOM in (\ref{eq 2.2}). 
The perturbation term proportional to $(\mathbf {x}-\mathbf {x}_c)$ in (\ref{eq 2.3}) is responsible for the appearance of the Berry curvature in $\dot{ \langle \mathbf {x} \rangle}$. Relations 
(\ref{eq 2.4}) are considered more complete as compared to (\ref{eq 2.2}) since they account for geometrical features associated with the Berry curvature of the single band responsible for the transport in a given material. 
The relations in (\ref{eq 2.4}) have also been obtained by a Hamiltonian approach \cite{Shindou-2005} which uses a projection procedure of operators onto a wave packet subspace consisting of a single energy band. 
The resulting operators from this projection lead to noncanonical commutation relations giving rise to the Berry phase contributions in $\dot{ \langle \mathbf {x} \rangle}$. 
Another approach \cite{Gosselin}, based on a semiclassical expansion and diagonalization  of the electromagnetic Hamiltonian, has also resulted in the same EOM (in (\ref{eq 2.4})). 
This method also relies on projected operators leading to noncanonical commutation relations \cite{Shindou-2005,Gosselin,Basu}.

These recent developments have shown that to take into account features from the Berry curvature in the EOM, one must carefully consider the slow time dynamics of the electromagnetic fields in the Hamiltonian. 
Another key issue is the application of the operator projection procedure as means to capture the role of a finite subset of bands that determines the transport in the solid. 
Interestingly, a rigorous derivation of such a projection has not been given yet, although some justification within a semiclasscal expansion has been offered in \cite{Panati}. 
Also, the majority of the works have focused on a single band projection with the exception of \cite{Culcer,Shindou-2008} where a finite set of identical bands of a multiband  packet was considered. We note that the EOM in (\ref{eq 2.4}) constitute a significant step towards a more complete understanding of transport in materials, however the outstanding questions of justifying the projection procedure and explicitly taking into account multiband wave packet with different dispersions must be resolved, especially for situations when the Fermi level is in close proximity to crossing points between bands. 

\begin{figure}
\includegraphics[width=0.22\textwidth]{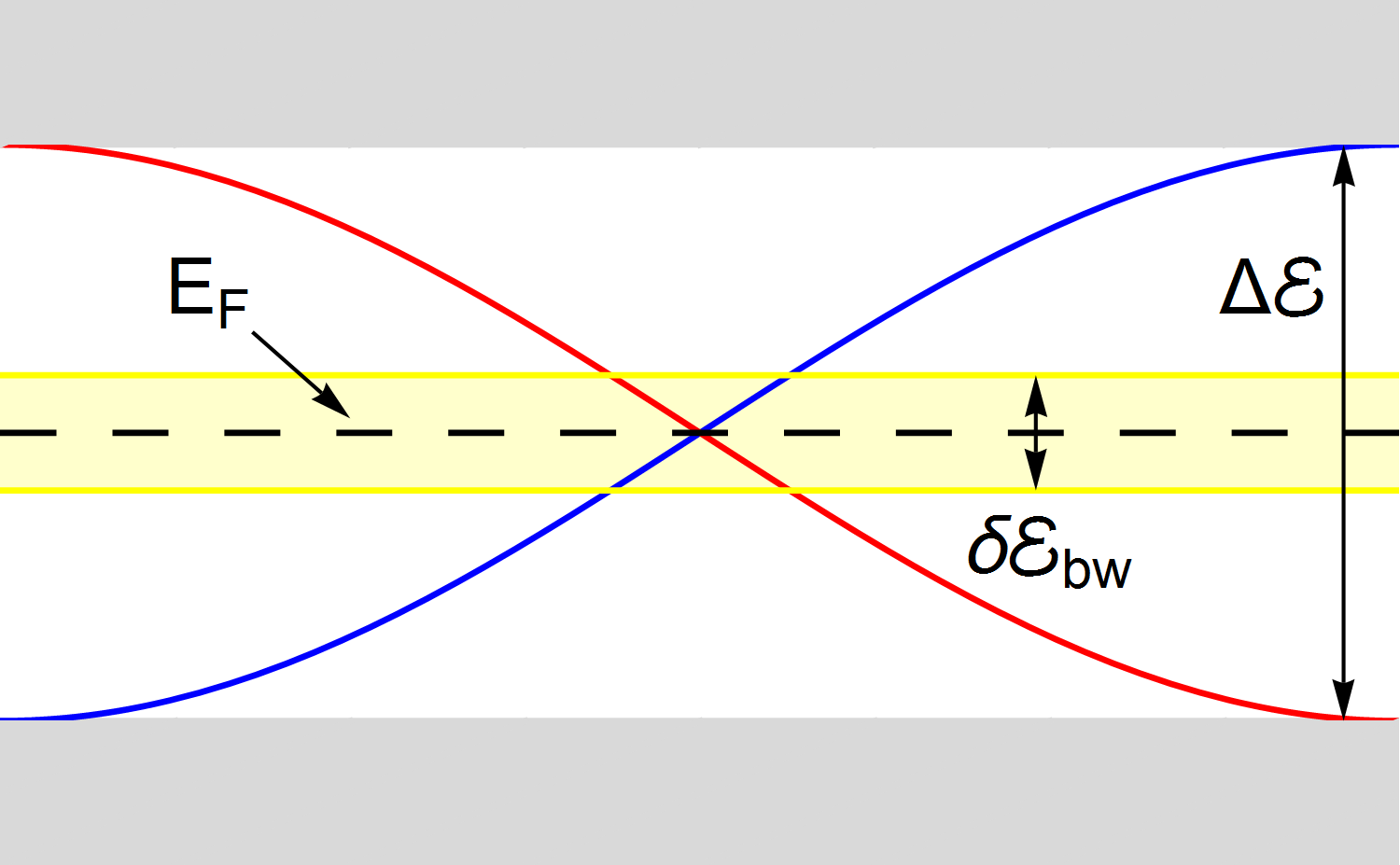}
\caption{Schematics of the $D$ energy band region with a band width $\delta \mathcal{E}_{bw}$ (yellow) separated from the rest of the band structure (gray) with a large gap $\Delta \mathcal{E}$, such that $\delta \mathcal{E}_{bw} \ll \Delta \mathcal{E}$. As an example, two linear bands  in the $D$ region with a crossing point in the vicinity of the Fermi level $E_F$ are shown.}
\label{fig2}
\end{figure}

{\it Projected EOM - } 
We first show that the projection procedure of operators follows naturally from perturbation theory and an adiabatic approximation. 
For this purpose, we  consider the following conditions: (i) there is a Hamiltonian of the form $H = H^0(\alpha(t)) + H^1(\alpha(t))$ where the unperturbed part $H^0$ may be time-dependent due to a set of time-dependent parameters, collectively denoted as $\alpha(t)$, that may also be present in the perturbation $H^1(\alpha(t))$ with perturbation parameter $\epsilon$, (ii) $\alpha(t)$ is turned on and thereafter varies slowly in time, and (iii) the eigenfunctions of $H^0(\alpha(t))$ are ${\psi_n^0(\alpha(t))}$ and are well separated from the other eigenstates by a large energy gap for all time. 
The subspace of these states is denoted as $D$ and indexed by $I_D$, with $\psi_n^0 \in D$ and $n \in I_D$. By applying perturbation theory in tandem with adiabaticity under these circumstances, the time dynamics of an observable $O$ is found to have the following form,
\begin{subequations} \label{eq 3.1}
\begin{align}
\dot{\langle O \rangle} &= \frac{i}{\hbar} \langle \left[ \mathcal{P}_D (H) , \mathcal{P}_D (O) \right] \rangle + \langle \frac{d}{dt}  \mathcal{P}_D (O) \rangle , \\
\mathcal{P}_D (O) &= \sum_{m,n \in I_D} | \psi_m^0 \rangle \langle \psi_m^0 | O | \psi_n^0 \rangle \langle \psi_n^0 |, \\
\dot c_n &= - \frac{i}{\hbar} c_n \langle \psi_n^0 | H | \psi_n^0 \rangle -  \sum_{m \in I_D} c_m \langle \psi_n^0 | \dot \psi_m^0 \rangle,
\end{align}
\end{subequations}
where $\mathcal{P}_D$ is the projection operator onto $D$.  
Here $c_n$ are the coefficients for the expansion of the wave packet $\psi = \sum_{n \in I_D} c_n \psi_n^0$ and their coupled dynamics is essentially given from the Schr{\"o}dinger equation restricted to the subspace $D$. We emphasize that the projection operator is a natural consequence of the assumptions for the adiabatic approximation and the large energy gap separation condition as described above. Extensive details of the derivation for the dynamics of $O$ and the expansion coefficients  $c_n$ are shown in the Supplemental Information. 

We wish to apply the projected dynamics in (\ref{eq 3.1}) to the Hamiltonian in (\ref{eq 2.3}) for a multiband wave packet centered in reciprocal space about wave vectors with energies in the vicinity of the Fermi level.
In this situation, the subspace $D$ consists of energy band states in the multiband wave packet near the Fermi level, which may also include crossing points. 
In the context of this perturbation theory with adiabaticity, the band width of the energy band states in $D$ is $\delta \mathcal{E}_{bw}$, which is separated by a large energy gap $\Delta \mathcal{E}$ from the rest of the band structure ($\delta \mathcal{E}_{bw}\ll \Delta \mathcal{E}$), as schematically illustrated in Fig. \ref{fig2}.

As explained in \cite{Niu.99} and the Supplementary Information, the series expansion of the electromagnetic Hamiltonian in (\ref{eq 2.3}) can be considered as having an unperturbed term coming from the bare crystal potential $H^0=H_c(\mathbf{p} - e\mathbf{A}_c , \mathbf{x} ; \mathbf{x}_c)$ and a perturbation $H^1=\frac{1}{2}\frac{\partial H_c}{\partial \mathbf{x}_c} \cdot (\mathbf{x} - \mathbf{x}_c) +h.c.$, associated with the derivatives of the electromagnetic potentials $\mathbf{A}_c$ and $\varphi_c$ with respect to $\mathbf{x}_c$. 
 The potentials $\mathbf{A}_c$ and $\varphi_c$ serve as the collective parameter $\alpha(t)$ in conditions (i)-(iii).
In addition to the perturbation as described, the slow time scale of these potentials meets conditions (i) and (ii).
Furthermore, in accordance with the wave packet description, the initial wave function is a multiband packet consisting of Bloch states as shown in $\left( \ref{eq 2.1} \right)$. 
The band width $\delta \mathcal{E}_{bw}$ of states near a Fermi level tuned to a crossing point is well separated from the rest of the band structure ($\delta \mathcal{E}_{bw} \ll \Delta \mathcal{E}$), leading to condition (iii). 
In Fig. \ref{fig2}, we schematically show this situation of two crossing bands with linear dispersion in the vicinity of the Fermi level. Given that conditions (i)-(iii) are met, the EOM for $\mathbf {x}$ and $ \mathbf {k}$ subject to the perturbative Hamiltonian in (\ref{eq 2.3}) are obtained using (\ref{eq 3.1}) in what follows for multiband packets.

{\it Results for the EOM - } Relations (\ref{eq 3.1}) provide a rather straight forward (although tedious) path to obtain the projected EOM for an observable provided the Hamiltonian for the system is given explicitly. 
The problem essentially involves working out several commutators. 
Here (\ref{eq 3.1}) is applied to obtain the EOM for $\mathbf {x}$ and $ \mathbf {k}$ for the electromagnetic Hamiltonian in its linearized form (in (\ref{eq 2.3})), using the eigenstates $\psi_n^0(\mathbf{q}) = e^{i\mathbf{q}\cdot\mathbf{x}}u_n^0(\mathbf{q}-e\mathbf{A}_c)$. 
For clarity and concreteness we take the circular gauge with $\mathbf{A} = \frac{1}{2} \left( \mathbf{B} \times \mathbf{x} \right)$ and $\varphi = - \mathbf{E} \cdot \mathbf{x}$, where the fields may be slowly varying in time. After some calculations (details are given in the Supplementary Information), we find 
\begin{widetext}
\begin{subequations}\label{eq 4.1}
\begin{align}
\dot{ \langle \mathbf {x} \rangle} &=  \langle \mathbf {v} \rangle
	- \frac{e}{2\hbar^2} |z_{n_1}|^2 \left\{ \left[ \left( \mathbf{x}_c -  \mathbf{\nu}_{\gamma_{n_1}} \right) \times \mathbf{B} \right]\cdot  \nabla \right\} \mathbf{v}_{n_1}
	 - \frac{e}{\hbar} z_{n_1}^*  z_{n_2} \bigg\{
	\frac{1}{4} \left[ \left( \mathbf{B} \times \bar{\mathbf{\mathcal{A}}}_{n_1 n_2} \right) 
	\cdot  \nabla \right] \left( \mathbf{v}_{n_1} + \mathbf{v}_{n_2} \right) 	\nonumber\\ 
&+ \frac{i}{4} \left[ \mathbf{B} \cdot \left( \left(\mathbf{\nu}_{\gamma_{n_1}} +\mathbf{\nu}_{\gamma_{n_2}}- \mathbf{x}_c\right) \times \bar{\mathbf{\mathcal{A}}}_{n_1 n_2} \right) \right]  \left( \mathbf{v}_{n_1}-  \mathbf{v}_{n_2} \right) 
+ \frac{i}{4} \left[ \mathbf{B} \cdot \left( \left( 
	\mathbf{\nu}_{\gamma_{n_1}} +\mathbf{\nu}_{\gamma_{n_2}} - \mathbf{x}_c \right) 
	\times \left( \mathbf{v}_{n_1}-  \mathbf{v}_{n_2} \right) \right) \right]   \bar{\mathbf{\mathcal{A}}}_{n_1 n_2}	 \nonumber \\
&+ \frac{1}{4} \left[ \left( \mathbf{v}_{n_1} + \mathbf{v}_{n_2} \right) \times \mathbf{B} \right] \times \bar{\mathbf{\Omega}}_{n_1 n_2} 
	- \frac{1}{4} \left( \mathbf{v}_{n_3} \times  \mathbf{B} \right) \times i \bar{\mathbf{Q}}_{n_1 n_3 n_2}
	+ \left[ \frac{1}{2} \left( \langle \mathbf {v} \rangle \times \mathbf{B} \right) + \mathbf{E} \right] 
	\times \left( \bar{\mathbf{\Omega}}_{n_1 n_2} -i \bar{\mathbf{Q}}_{n_1 n_3 n_2} \right)  \nonumber \\
&+  \frac{1}{e}  \nabla \left( \mathbf{B} \cdot \bar{\mathbf{M}}_{n_1 n_2} \right)
	+ i \mathbf{B} \cdot \left( \frac{1}{4} \bar{\mathbf{\mathcal{A}}}_{n_1 n_3} \times \mathbf{v}_{n_1} 
	- \frac{1}{e} \bar{\mathbf{M}}_{n_1 n_3} \right) 
	 \bar{\mathbf{\mathcal{A}}}_{n_3 n_2}
	- i \mathbf{B} \cdot \left( \frac{1}{4} \bar{\mathbf{\mathcal{A}}}_{n_3 n_2} \times \mathbf{v}_{n_2} 
	- \frac{1}{e} \bar{\mathbf{M}}_{n_3 n_2} \right) 
	 \bar{\mathbf{\mathcal{A}}}_{n_1 n_3} \bigg\},    \\
\dot{ \langle \mathbf {k} \rangle} &= \frac{e}{\hbar}\left(\mathbf E+ \langle \mathbf {v} \rangle \times \mathbf B\right),     \\
i \hbar \dot{z}_n &= \left\{ \mathcal{E}_n + e \left[ \frac{1}{2} \left( \mathbf{v}_n \times \mathbf{B} \right) + \mathbf{E} \right] \cdot \mathbf{x}_c \right\} z_n
- \sum_{m \in I_D}\left\{ \frac{e}{2} \left[ \left( \frac{1}{2} \left( \mathbf{v}_n + \mathbf{v}_m \right) + \dot{ \langle \mathbf {x} \rangle} \right) \times \mathbf{B} + 2 \mathbf{E} \right] \cdot \bar{\mathcal{A}}_{nm} + \mathbf{B} \cdot \bar{\mathbf{M}}_{nm} \right\} z_m    \nonumber \\
&- \frac{ie}{4} \left[ \left( \mathbf{v}_n - \langle \mathbf {v} \rangle \right) \times \mathbf{B} \right] \cdot \nabla z_n,
\end{align}
\end{subequations}
\end{widetext}
where summation over all bands in the multiband wave packet for each band index $n_1, n_2,$ and $n_3$ in each term in $\dot{ \langle \mathbf {x} \rangle}$ and $\dot{ \langle \mathbf {k} \rangle}$ is implied. 
We have defined the interband Berry connection $ \bar {\mathbf{\mathcal{A}}}_{n_1 n_2} = i \langle u_{n_1} (\mathbf{k}_c) |   \nabla | u_{n_2} (\mathbf{k}_c) \rangle$ and interband Berry curvature $\bar {\mathbf{\Omega}}_{n_1 n_2} = \nabla \times \bar{\mathbf{\mathcal{A}}}_{n_1 n_2}$, as well as an interband magnetization $ \bar{\mathbf{M}}_{n_1 n_2}= - \frac{ie}{4\hbar} \big<  \nabla u_{n_1} (\mathbf{k}_c)| 
\left( \mathcal{E}_{n_1}(\mathbf{k}_c) + \mathcal{E}_{n_2}(\mathbf{k}_c) - 2H_0 (\mathbf{p}+\hbar\mathbf{k}_c,\mathbf{x})\right)$ $\times |  \nabla  u_{n_2}(\mathbf{k}_c) \big> $. 
The non-Abelian coupling between interband Berry connections is denoted as 
$ \bar{\mathbf{Q}}_{n_1 n_2 n_3}=\bar{\mathbf{\mathcal{A}}}_{n_1 n_2} \times \bar{\mathbf{\mathcal{A}}}_{n_2 n_3} $. 
We note that $\mathbf{x}_c$ and $\mathbf{k}_c$ are expectation values calculated with the wave function $\psi = \sum_{n\in I_D} \int d^3\mathbf{k} c_n(\mathbf {k})\psi_n^0(\mathbf {k}) e^{- \frac{i}{\hbar} \int_{0}^{t} E_n^0(\mathbf{k})(t') dt'}$ for the wave packet. 
The coefficients $c_n$ are found from first solving for the $z_n$ in (\ref{eq 4.1}) and then using $\left( \ref{eq 2.1} \right)$.
Also, $ \langle \mathbf {v} \rangle=\sum_{n \in I_D}a_n^2 \mathbf{v}_n$ depends only on the wave packet distribution across bands and their dispersions, 
$\mathbf{v}_n=\frac{1}{\hbar} \nabla \mathcal{E}_n$, $\mathbf{\nu}_{\gamma_n}=\nabla \gamma_n$, and $\nabla$ is understood as differentiation with respect to $\mathbf{k}_c$ everywhere.

A comparison of the above results  with the singleband EOM in (\ref{eq 2.4}) shows that the results for $ \dot{ \langle \mathbf {x} \rangle}$ are much more complicated for the multiband packet. 
The first term in (\ref{eq 4.1}a) is simply the standard group velocity weighted over all bands. 
The remaining terms on the first line and all terms on the second line capture various aspects of the different band dispersions. 
In the singleband case, these dispersion terms reduce to the $\nabla \mathcal{E}(\mathbf {k}_c)$ contribution in $ \dot{ \langle \mathbf {x} \rangle}$ in (\ref{eq 2.4}). 
All terms in the third line and the terms containing $\mathbf v_n$ in the fourth line correspond to the anomalous velocity, which now reflects the different dispersions and interband Berry curvatures and connections.
Clearly, this is more complicated when comparing with the singleband $\dot{ \langle \mathbf {k} \rangle} \times \bar {\mathbf{\Omega}}_n$ anomalous velocity in (\ref{eq 2.4}). 
The rest of the terms on the last line consist of energy contributions of the magnetic field coupled to the interband magnetizations, which corresponds to the $\bar {\mathbf{M}}_n \cdot \mathbf{B} $ contribution to $ \dot{ \langle \mathbf {x} \rangle}$ in (\ref{eq 2.4}). 
Surprisingly, the only difference between the $\dot{ \langle \mathbf {k} \rangle}$ in (\ref{eq 4.1}) and that in (\ref{eq 2.4}) is the  $ \langle \mathbf {v} \rangle$ that appears in (\ref{eq 4.1}) instead of the full expression for $\dot{ \langle \mathbf {x} \rangle}$. 
The reason that  $\dot{ \langle \mathbf {k} \rangle}$ in (\ref{eq 4.1}) is similar to $\dot{ \langle \mathbf {k} \rangle}$ in (\ref{eq 2.4}) is because  $\mathbf{k}$ is diagonal in the zeroth order Bloch basis. 
Physically, the real space wave packet doesn't "see" the distribution across bands in reciprocal space in regards to the momentum and we thus retain a Lorentz-like force expression for $\dot{ \langle \mathbf {k} \rangle}$.

We examine (\ref{eq 4.1}) in some specific cases. 
Firstly, the single band wave packet EOM reduce to $\dot{ \langle \mathbf {x} \rangle}=\frac{1}{\hbar}\nabla \mathcal{E}_M (\mathbf {k}_c) - \dot{ \langle \mathbf {k} \rangle} \times \bar{\mathbf{\Omega}}$ and 
$\dot{ \langle \mathbf {k} \rangle} = \frac{e}{\hbar}(\mathbf E+ \langle \mathbf {v} \rangle \times \mathbf B)$,
which is the same as (\ref{eq 2.4}) except for the $ \langle \mathbf {v} \rangle$ that appears in $\dot{ \langle \mathbf {k} \rangle}$ here as opposed to  $\dot{ \langle \mathbf {x} \rangle}$. We also consider the case of degenerate bands, i.e. $\mathcal{E}_n=\mathcal{E}$ for all $N$ bands, for which (\ref{eq 4.1}) reduces to 
\begin{widetext}
\begin{subequations}\label{eq 4.2}
\begin{align}
\dot{ \langle \mathbf {x} \rangle}&=\langle \mathbf {v} \rangle  
	- \frac{1}{\hbar} z_{n_1}^*  z_{n_2} \Biggl[ 
	\nabla \left( \mathbf{B} \cdot \bar{\mathbf{M}}_{n_1 n_2} \right) 
	- i \mathbf{B} \cdot \bar{\mathbf{M}}_{n_1 n_3} \bar{\mathbf{\mathcal{A}}}_{n_3 n_2}
	+ i \mathbf{B} \cdot \bar{\mathbf{M}}_{n_3 n_2} \bar{\mathbf{\mathcal{A}}}_{n_1 n_3}   
	- \dot{ \langle \mathbf {k} \rangle}\times \left( \bar{\mathbf{\Omega}}_{n_1 n_2} - i \bar{\mathbf{Q}}_{n_1 n_3 n_2} \right) \Biggr],   \\
\dot{ \langle \mathbf {k} \rangle}&= \frac{e}{\hbar}\left(\mathbf E+ \langle \mathbf {v} \rangle \times \mathbf B\right),    \\
i \hbar \dot{z}_n &= \left\{ \mathcal{E} + e \left[ \frac{1}{2\hbar} \left( \nabla \mathcal{E} \times \mathbf{B} \right) + \mathbf{E} \right] \cdot \mathbf{x}_c \right\} z_n   
- \sum_{m \in I_D} \biggl\{ \frac{e}{2} \left[ \left( \frac{1}{\hbar} \nabla \mathcal{E} + \dot{ \langle \mathbf {x} \rangle} \right) \times \mathbf{B} + 2 \mathbf{E} \right] \cdot \bar{\mathcal{A}}_{nm}  
+ \mathbf{B} \cdot \bar{\mathbf{M}}_{nm} \biggl\} z_m.
\end{align}
\end{subequations}
\end{widetext}
All notation and conventions for summation over band indices and $\nabla$ differentiation are the same as in (\ref{eq 4.1}). It is noted that the above results are of the same form as in  \cite{Shindou-2005}, where the wave packet localized in $N$ degenerate bands was also considered. 
Comparing the results in (\ref{eq 4.2}a) with the general expression in (\ref{eq 4.1}a) shows that the only terms that do not vanish in $\dot{ \langle \mathbf {x} \rangle}$ are those coming from the coupling between the magnetic field and interband magnetizations. 
Also, the anomalous velocity now contains only terms coupling $\dot{ \langle \mathbf {k} \rangle}$  to $\bar{\mathbf{\Omega}}_{n_1 n_2}$ and $\bar{\mathbf{Q}}_{n_1 n_3 n_2}$, which are manifestations of the non-canonical commutation relations of projected position operators onto the multiband subspace.

Finally, (\ref{eq 4.1}) is applied to the case of a pair of linear bands with dispersion $\mathcal{E}_{\pm} (\mathbf {k})= \pm v k$, where $v$ is the isotropic Fermi velocity (schematics in Fig. 2). 
Such a band structure corresponds to a Hamiltonian $H = v \boldsymbol{\sigma} \cdot \mathbf {k}$ ($\boldsymbol{\sigma}$ are the Pauli matrices) in the vicinity of a linear two band crossing in a Weyl semimetal or graphene, for example \cite{Wehling,RevModPhys-2018}. 
 By using the corresponding eigenstates $\psi_+ = \begin{pmatrix} \cos\frac{\theta}{2} \\ e^{i\varphi}\sin\frac{\theta}{2} \end{pmatrix}$ and $\psi_- = \begin{pmatrix} e^{-i\varphi}\sin\frac{\theta}{2} \\ -\cos\frac{\theta}{2} \end{pmatrix}$ of this $H$ 
as local forms for the Bloch functions, one obtains
\begin{widetext}
\begin{subequations}\label{eq 4.3}
\begin{align}
\dot{ \langle \mathbf {x} \rangle}
	&= \langle \mathbf {v} \rangle - \frac{1}{\hbar} |z_{n_1}|^2  \nabla \left( \mathbf{B} \cdot \bar{\mathbf{M}}_{n_1 n_1}  \right)
	- \frac{e}{\hbar^2}  z_{n_1}^*  z_{n_2} \bigg\{
       	\frac{i}{4} \left[ \mathbf{B} \cdot \left( \left( 
	\mathbf{\nu}_{\gamma_{n_1}} +\mathbf{\nu}_{\gamma_{n_2}} - \mathbf{x}_c \right) 
	\times \left( \mathbf{v}_{n_1} - \mathbf{v}_{n_2} \right) \right) \right] \bar{\mathbf{\mathcal{A}}}_{n_1 n_2}  \nonumber  \\
	&+ \frac{i}{4} \left[ \mathbf{B} \cdot \left( \left( 
	\mathbf{\nu}_{\gamma_{n_1}} +\mathbf{\nu}_{\gamma_{n_2}} - \mathbf{x}_c \right) 
	\times \bar{\mathbf{\mathcal{A}}}_{n_1 n_2} \right) \right]  \left( \mathbf{v}_{n_1} - \mathbf{v}_{n_2} \right) 
	+ \frac{1}{4} \left[ \left( \mathbf{v}_{n_1} + \mathbf{v}_{n_2} \right) \times \mathbf{B} \right] \times \bar{\mathbf{\Omega}}_{n_1 n_2}
		            \bigg\},   \\
	\dot{ \langle \mathbf {k} \rangle} &= \frac{e}{\hbar}\left(\mathbf E+ \langle \mathbf {v} \rangle \times \mathbf B\right),    \\
i \hbar \dot{z}_n &= \left[ \mathcal{E}_n + e \left( \frac{1}{2\hbar} \left( \nabla \mathcal{E}_n \times \mathbf{B} \right) + \mathbf{E} \right) \cdot \mathbf{x}_c - \frac{e}{2\hbar} \left( \nabla \mathcal{E}_n \times \mathbf{B} \right) \cdot \bar{\mathcal{A}}_{nn} \right] z_n
	- \sum_{m \in \pm}\left[ \frac{e}{2} \left( \dot{ \langle \mathbf {x} \rangle} \times \mathbf{B} + 2 \mathbf{E} \right) \cdot \bar{\mathcal{A}}_{nm} + \mathbf{B} \cdot \bar{\mathbf{M}}_{nm} \right] z_m    	\nonumber \\
	&- \frac{ie}{4} \left[ \left( \mathbf{v}_n - \langle \mathbf {v} \rangle \right) \times \mathbf{B} \right] \cdot \nabla z_n,
\end{align}
\end{subequations}
\end{widetext}
where $\{n_1, n_2\}=\pm$. Here $\langle \mathbf {v} \rangle=\frac{v}{\hbar} \left( a_+^2 - a_-^2 \right)\hat{k}$, which suggests that this term would vanish if the distribution across the two bands is equal. Because of the linear dispersion many  terms in the general case in (\ref{eq 4.1}a) vanish or become modified. 
Specifically, there is only one term coming from the coupling between the interband magnetization, while dispersion terms with $\mathbf {v}_{n_1}-\mathbf {v}_{n_2}$ account for differences in energy band slopes. 
The anomalous velocity (last term in (\ref{eq 4.3}a)) is now associated with the interband Berry curvature and the band dispersions. 
We note that there is an explicit dependence on $a_\pm$, $\gamma_\pm$ in the exponents and in the $\mathbf{\nu}_{\gamma_\pm}$, thus to obtain more specific results, detailed knowledge of the wave packet properties is needed, which is beyond the scope of this paper. 


{\it Conclusions - }In this study, the time dynamics of the position and gauge invariant crystal momentum subject to electromagnetic fields in a periodic environment is considered using a general Hamiltonian constructed for a localized multiband wave packet. 
One of the main contributions here is showing that when the wave packet is spread over a subset of bands well-separated from the rest of the band structure of the crystal, a projection procedure of the Hamiltonian and EOM can be applied. 
We demonstrate that such a procedure follows naturally from an adiabatic approximation within perturbation theory, whereas this justification is mostly lacking in previous works. 
This projection procedure reflects that transport is dominated by a finite number of energy bands in the vicinity of the Fermi level. 
Another merit of our work is that this Hamiltonian-based approach is rather transparent and accounts for different energy band dispersions that make up the wave packet. 
Perhaps the most noteworthy result comes from the found interband effects in the anomalous velocity. 
These are captured via non-Abelian Berry characteristics giving new perspectives for nontrivial topology. 
The application of this general theory to several cases (including two linear crossing bands) shows that the generalized EOM depend strongly on the explicit band structure of the material and quantum mechanical effects beyond the single band wave packet description can lead to novel properties. 

To further understand how specific features in the EOM for a multiband wave packet affect transport however, one must utilize a multiband generalization of the Boltzmann equation \cite{Ashcroft} in order to obtain specific transport properties, such as electric, heat, and spin currents. 
The EOM described in this work constitute much needed ground for such a generalized transport approach that will enable finding transport currents, Hall effects and other properties influenced by interband properties like the non-Abelian Berry curvature, which could lead to measurable signatures in the laboratory.

We are thankful to Prof. Qian Niu for discussions. L.W. acknowledges financial support from the US Department of Energy under grant No. DE-FG02-06ER46297. C.T. acknowledges financial support from the Deutsche Forschungsgemeinschaft, in part through Research Training Group GRK 1621 and Collaborative Research Center SFB 1143, project A04. T.S. also acknowledges support from the ERASMUS+ Programme of the European Union and the hospitality of TU Dresden.

\end{document}


\title{Supplementary Information: Multiband Effects in Equations of Motion of Observables Beyond the Semiclassical Approach}

\author{Troy Stedman}
\affiliation{Department of Physics, University of South Florida, Tampa, Florida 33620, USA}

\author{Carsten Timm}
\affiliation{Institute of Theoretical Physics and Center for Transport 
and Devices of Emergent Materials, Technische Universit{\"a}t Dresden, 
01062 Dresden, Germany}

\author{Lilia M. Woods}
\affiliation{Department of Physics, University of South Florida, Tampa, Florida 33620, USA}
\date{\today}





\date{\today}

\maketitle


\section{Adiabatic Approximation and Perturbation Theory}

Here we demonstrate in detail how the equations of motion (EOM) projected onto a subset of locally degenerate or nearly degenerate bands arise from perturbation theory with an adiabatic condition ((5) and (6) from the main text). We begin with a general situation, which will be related to the situation in the main text.

{\it General Model:} We consider a general Hamiltonian given as $H(\alpha(t))=H^0 (\alpha(t))+H^1 (\alpha(t))$, where the time dependence is characterized by the composite parameter $\alpha (t)$. 
$H^0$ has a degenerate spectrum with degenerate subspace $D$ and the perturbation $H^1 (\alpha(t))$, is turned on at time $t=0$. 
The eigenstates for the full Hamiltonian are the set $\{\psi_n\}$ with $H\psi_n(\alpha(t)) = E_n(\alpha(t)) \psi_n(\alpha(t))$ and the eigenstates for the unperturbed Hamiltonian are the set $\{\psi_n^0\}$ with $H^0 \psi_n^0 (\alpha(t))= E_n^0(\alpha(t)) \psi_n^0(\alpha(t))$. 
For the following, we utilize states $\psi_n^0 \in D$ that diagonalize $H^1$ in $D$. 
We require a large energy gap condition on $D$, meaning that the energies of eigenstates in $D$ are separated from all other energies by a large energy gap.  We also assume that a particle starts out in the degenerate subspace $D$ of $H^0$ at $t=0$ and the large energy gap condition is maintained for the instantaneous energies of $H^0$ while $D$ changes in time. 

{\it Perturbation:} It is noted that the Hamiltonian in the derivation of the EOM of operators in the main text involves the perturbation $H^1 = \frac{1}{2}\frac{\partial H_c}{\partial \mathbf{x}_c} \cdot (\mathbf{x} - \mathbf{x}_c) +h.c.$, where the perturbation parameter $\epsilon$ is $\epsilon = \frac{\Delta x}{\lambda}$ with $\Delta x$ as the wave packet spread in real space and $\lambda$ as the wavelength of the Fourier components of the vector potential. 
It's clear that $\Delta x$ comes from the operator $(\mathbf{x} - \mathbf{x}_c)$. 
To show that $\frac{\partial H_c}{\partial \mathbf{x}_c}$ is proportional to $\frac{1}{\lambda}$, we first simply observe that $\frac{\partial H_c}{\partial \mathbf{x}_c} = \sum_j \frac{\partial H_c}{\partial A_{c,j}} \frac{\partial A_{c,j}}{\partial \mathbf{x}_c}$. 
Now we write $\mathbf{A}_c$ as a Fourier transform, 

\begin{myeqnarray}\label{eq1.1}
\mathbf{A}_c = \int \tilde{\mathbf{A}}(\mathbf{q}) e^{-i\mathbf{q} \cdot \mathbf{x_c}} d^3 \mathbf{q}.
\end{myeqnarray}
Here $q = \frac{2 \pi}{\lambda}$ is the magnitude of the Fourier wave vector component with corresponding wavelength $\lambda$ and $\mathbf{x_c}$ is the location of the center of the wave packet in real space. 
In the long-wavelength limit, only those $\mathbf{q}$ with large $\lambda$ contribute appreciably to the Fourier transform. 
Therefore, $\tilde{\mathbf{A}}(\mathbf{q})$ is non-negligible only in the vicinity of $\mathbf{q}\approx0$. 
Additionally, one finds for the spatial derivative $\frac{\partial A_{c,j}}{\partial x_{c,k}} = i \int q_k \tilde{A}_j(\mathbf{q}) e^{-i\mathbf{q} \cdot \mathbf{x_c}} d^3 \mathbf{q} \propto \frac{1}{\lambda}$.

{\it Expansion Coefficients:} Our goal is to study the dynamics of Bloch wave packets in the vicinity of crossing points in the band structure for the linearized Hamiltonian in the main text (Eq. (3)). 
If the Fermi level of our system is tuned to such a crossing point, then the Bloch wave packet consists of these degenerate states. 
The degenerate states gives rise to the degenerate subspace $D$ that is separated from all other band states by a large energy gap. 
 Therefore, the goal to describe the dynamics of such Bloch wave packets meets the requirements of the general approach outlined above. 

Returning to the general situation, we start by expanding the full wave function $\psi(t)$ for the particle in terms of the full Hamiltonian's instantaneous eigenstates $\psi_n (t)$,  
\begin{myeqnarray}\label{eq1.2}
\psi (t) = \sum_{n \in I_\mathcal{H}} c_n (t) \psi_n (t),
\end{myeqnarray}
where the sum is over the index set $I_{\mathcal{H}}$ for a complete set of instantaneous eigenstates in the Hilbert space $\mathcal{H}$.

For a time-dependent Hamiltonian, one can write the Schr{\"o}dinger equation for the expansion coefficients
\begin{myeqnarray}\label{eq1.3}
\dot c_n = - \frac{i}{\hbar} c_n E_n -  \sum_{m \in I_\mathcal{H}} c_m \langle \psi_n | \dot \psi_m \rangle.
\end{myeqnarray}

The instantaneous eigenstates $\psi_m$ are perturbatively expanded as
\begin{myeqnarray}\label{eq1.4}
\psi_m = \psi_m^0 + \sum_{n \in I_\mathcal{H}} \left( c_{m,n}^1 + c_{m,n}^2 + ... \right) \psi_n^0.
\end{myeqnarray}

One notes that each coefficient $c_{m,n}^j$ for $j>0$ contains factors of the form $\frac{\langle \psi_k^0 | H^1 | \psi_l^0 \rangle}{E_l^0 - E_k^0}$ for some states $\psi_k^0 , \psi_l^0$ and energies $ E_k^0 ,  E_l^0$. 
Since states in $D$ are separated from all other states by a large energy gap, such terms are negligible when either 
$\psi_k^0 \notin D$ and $\psi_l^0 \in D$ or vice versa, and thus (using the index set $I_D$ for the set of basis eigenstates that span $D$)
\begin{myeqnarray}\label{eq1.5}
\psi_m =
\begin{cases}
\psi_m^0 + \sum_{n \notin D} \left( c_{m,n}^1 + c_{m,n}^2 + ... \right) \psi_n^0, & m \notin I_D
\\ \psi_m^0, & m \in I_D.
\end{cases}
\end{myeqnarray}

The above result implies that $\psi_m$ is approximated as simply $\psi_m^0$ for $m \in I_D$ and is entirely contained outside of $D$ when $m \notin I_D$. 
Next, the $k^{th}$ order of the expansion coefficients from (\ref{eq1.3}) are considered and can be cast as 
\begin{myeqnarray}\label{eq1.6}
\dot c_n^k = - \frac{i}{\hbar} \sum_{j=0}^k c_n^{k-j} E_n^j - \sum_{j=0}^{k-i} \sum_{i=0}^{k} \sum_{m \in I_\mathcal{H}} c_m^{k-i-j} \langle \psi_n^j | \dot \psi_m^i \rangle.
\end{myeqnarray}

It is realized that in the zeroth order term, $\dot c_n^0 = - \frac{i}{\hbar} c_n^0 E_n^0 - \sum_{m \in I_\mathcal{H}} c_m^0 \langle \psi_n^0 | \dot \psi_m^0 \rangle$, one can make use of the adiabatic approximation in the term containing $\langle \psi_n^0 | \dot \psi_m^0 \rangle$, giving
\begin{myeqnarray}\label{eq1.8}
\dot c_n^0 =
\begin{cases}
- c_n^0 \left( \frac{i}{\hbar} E_n^0 + \langle \psi_n^0 | \dot \psi_n^0 \rangle \right), & n \notin I_D
\\ - \frac{i}{\hbar} c_n^0 E_n^0 - \sum_{m \in I_D} c_m^0 \langle \psi_n^0 | \dot \psi_m^0 \rangle, & n \in I_D.
\end{cases}
\end{myeqnarray}

From the above relation it is realized that since there is no perturbation at $t=0$, then 
\begin{myeqnarray}\label{eq1.10}
c_n^0 = 0, n \notin I_D.
\end{myeqnarray}
Consequently, there is no zeroth order contribution to the coefficients $c_n$ for $n \notin I_D$ as expected from adiabatic theory. 
To zeroth order, the particle stays in the degenerate space $D$. 

We now claim that $c_n \approx 0$ to all orders for $n \notin I_D$, which can be shown by induction. 
Suppose that $c_n^j=0$ for all $j<k$ when $n \notin I_D$. 
Thus using (\ref{eq1.6}), for $n \notin I_D$, 
\begin{myeqnarray}\label{eq1.11}
\dot c_n^k = - \frac{i}{\hbar} c_n^0 E_n^0 - \sum_{m \notin I_D} c_m^k \langle \psi_n^0 | \dot \psi_m^0 \rangle - \sum_{j=0}^{k-i} \sum_{i=0}^{k} \sum_{m \in I_D} c_m^{k-i-j} \langle \psi_n^j | \dot \psi_m^i \rangle.
\end{myeqnarray}

By invoking the adiabatic approximation, the only term in the first sum that is non-negligible occurs when $m=n$. 
By invoking the large energy gap condition and (\ref{eq1.5}), the terms in the second sum are negligible. 
Therefore, the non-negligible terms give 
\begin{myeqnarray}\label{eq1.12}
\dot c_n^k = - c_n^k \left( \frac{i}{\hbar} E_n^0 + \langle \psi_n^0 | \dot \psi_n^0 \rangle \right) \rightarrow c_n^k = c_n^k (t=0) e^{- \frac{i}{\hbar} \int_{0}^{t} E_n^0(t') dt' - \int_{0}^{t} \langle \psi_n^0(t') | \dot \psi_n^0(t') \rangle dt'}, n \notin I_D.
\end{myeqnarray}

However, there is no is perturbation at $t=0$ and thus
\begin{myeqnarray}\label{eq1.13}
c_n^k = 0, n \notin I_D.
\end{myeqnarray}
Consequently, there is no $k^{th}$ order contribution to the coefficients $c_n$ for $n \notin I_D$.  
Coupled to the fact that $c_n^0 \approx 0$ for $n \notin I_D$, induction shows that $c_n = 0$ to all orders for $n \notin I_D$. 
Therefore, (\ref{eq1.3}) transforms to 
\begin{myeqnarray}\label{eq1.14}
\dot c_n = - \frac{i}{\hbar} c_n E_n -  \sum_{m \in I_D} c_m \langle \psi_n | \dot \psi_m \rangle,
\end{myeqnarray}
where the summation now runs over $I_D$ instead of $I_{\mathcal{H}}$. We are interested in (\ref{eq1.14}) when $n \in I_D$.  By using the large energy gap condition and (\ref{eq1.5}), (\ref{eq1.14}), it is found that 
\begin{myeqnarray}\label{eq1.15}
\dot c_n = - \frac{i}{\hbar} c_n E_n -  \sum_{m \in I_D} c_m \langle \psi_n^0 | \dot \psi_m^0 \rangle   ,   n \in I_D.
\end{myeqnarray}

One further notes that $\psi_n = \psi_n^0$ for $n \in I_D$ from (\ref{eq1.5}), thus the instantaneous energies can be written as
\begin{myeqnarray}\label{eq1.17}
H \psi_n^0 = E_n \psi_n^0   ,   n \in I_D. 
\end{myeqnarray}

We arrive at the important result for the expansion coefficients in (\ref{eq1.15}),
\begin{myeqnarray}\label{eq1.19}
\dot c_n = - \frac{i}{\hbar} c_n \langle \psi_n^0 | H | \psi_n^0 \rangle -  \sum_{m \in I_D} c_m \langle \psi_n^0 | \dot \psi_m^0 \rangle   ,   n \in I_D.
\end{myeqnarray}

{\it Dynamics of an Observable:} We now turn to the dynamics of an observable $O$,
\begin{myeqnarray}\label{eq1.20}
\dot{\langle O \rangle} = \sum_{m,n \in I_{\mathcal{H}}} \left( \dot c_m^* c_n + c_m^* \dot c_n \right ) \langle \psi_m | O | \psi_n \rangle +  c_m^*  c_n \frac{d}{dt} \langle \psi_m | O | \psi_n \rangle.
\end{myeqnarray}

Due to the approximation that $c_n = 0$ to all orders for $n \notin I_D$ and (\ref{eq1.5}), (\ref{eq1.20}) can be approximated as
\begin{myeqnarray}\label{eq1.21}
\dot{\langle O \rangle} = \sum_{m,n \in I_D} \left( \dot c_m^* c_n + c_m^* \dot c_n \right ) \langle \psi_m^0 | O | \psi_n^0 \rangle +  c_m^*  c_n \frac{d}{dt} \langle \psi_m^0 | O | \psi_n^0 \rangle.
\end{myeqnarray}

If we use the results in (\ref{eq1.19}), we find
\begin{myeqnarray}\label{eq1.22}
\dot{\langle O \rangle} &=& \frac{i}{\hbar} \sum_{m,n \in I_D} c_m^* c_n \left(  \langle \psi_m^0 | H | \psi_m^0 \rangle \langle \psi_m^0 | O | \psi_n^0 \rangle -  \langle \psi_m^0 | O | \psi_n^0 \rangle \langle \psi_n^0 | H | \psi_n^0 \rangle \right)  
		\nonumber \\ &+& \sum_{m,n \in I_D}  c_m^*  c_n  \frac{d}{dt} \langle \psi_m^0 | O | \psi_n^0 \rangle - \sum_{k,m,n \in I_D}  \left(  c_k \langle \psi_n^0 | \dot \psi_k^0 \rangle + c_k^* \langle \dot \psi_k^0 | \psi_m^0 \rangle \right) \langle \psi_m^0 | O | \psi_n^0 \rangle.
\end{myeqnarray}

Since $H = H^0 + H^1$ is diagonal in $D$ using the basis $\left\{ \psi_n^0 \right\}$, we can rewrite (\ref{eq1.22}) as 
\begin{myeqnarray}\label{eq1.23}
\dot{\langle O \rangle} &=& \frac{i}{\hbar} \sum_{k,m,n \in I_D} c_m^* c_n \left(  \langle \psi_m^0 | H | \psi_k^0 \rangle \langle \psi_k^0 | O | \psi_n^0 \rangle -  \langle \psi_m^0 | O | \psi_k^0 \rangle \langle \psi_k^0 | H | \psi_n^0 \rangle \right)  
		\nonumber \\ &+& \sum_{m,n \in I_D}  c_m^*  c_n  \frac{d}{dt} \langle \psi_m^0 | O | \psi_n^0 \rangle + \sum_{k,m,n \in I_D}  \left(  c_k \langle \dot \psi_n^0 | \psi_k^0 \rangle + c_k^* \langle \psi_k^0 | \dot \psi_m^0 \rangle \right) \langle \psi_m^0 | O | \psi_n^0 \rangle.
\end{myeqnarray}
Here we also moved the time derivatives on the states in the last two terms in (\ref{eq1.22}).

Finally, (\ref{eq1.23}) can be cast as 
\begin{mysubeqnarray}\label{eq1.24}
\begin{align}
\dot{\langle O \rangle} &= \frac{i}{\hbar} \left[ \mathcal{P}_D (H) , \mathcal{P}_D (O) \right] + \langle \frac{d}{dt}  \mathcal{P}_D (O) \rangle , \\
\mathcal{P}_D (O) &= \sum_{m,n \in I_D} | \psi_m^0 \rangle \langle \psi_m^0 | O | \psi_n^0 \rangle \langle \psi_n^0 | .
\end{align}
\end{mysubeqnarray}

These are the projected EOM with $\psi = \sum_{n \in I_D} c_n (t)\psi_n^0 $ and $c_n (t)$ are the solutions to (\ref{eq1.19}).
The essence of the projection of operators is that if a particle starts  out in the subspace $D$, then the EOM take a familiar Heisenberg form with operators projected onto $D$. 
It is this projection of operators that gives rise to (5) and (6) in the main text, leading to altered commutation relations for the operators time dynamics.

{\it Dynamics of the Coefficients:} We wish to write (\ref{eq1.19}) for the case of a Bloch wave packet in $D$, for which $ \psi = \sum_{n \in I_D} \int d^3\mathbf{q} c_n(\mathbf {q})\psi_n(\mathbf {q})$, with $\psi_n^0(\mathbf{q}) = e^{i\mathbf{q}\cdot\mathbf{x}}u_n^0(\mathbf{q}-e\mathbf{A}_c)$ and $c_n(\mathbf{q}) = \sqrt{\rho (\mathbf{q} - \mathbf{q}_c) } z_n (\mathbf{q}) $. 
With this form for the wave packet, (\ref{eq1.19}) becomes

\begin{myeqnarray}\label{eq4.1}
\dot c_n (\mathbf{k}) = - \frac{i}{\hbar} \sum_{m \in I_D} \int d^3 \mathbf{q} c_m (\mathbf{q}) \langle \psi_n^0 (\mathbf{k}) | H | \psi_m^0 (\mathbf{q}) \rangle -  \sum_{m \in I_D} c_m (\mathbf{k}) \langle \psi_n^0 (\mathbf{k}) | \dot \psi_m^0 (\mathbf{k}) \rangle   ,   n \in I_D.
\end{myeqnarray}

In addition to the diagonal unperturbed elements of the Hamiltonian, using (\ref{eq3.4}) leads to

\begin{myeqnarray}\label{eq4.2}
\sum_{m \in I_D} \int d^3 \mathbf{q} c_m (\mathbf{q}) \langle \psi_n^0 (\mathbf{k}) | H | \psi_m^0 (\mathbf{q}) \rangle &=& i \nabla_{\mathbf{x}_c} \mathcal{E}_n \cdot \nabla_{\mathbf{k}} c_n 
+ \frac{i}{2} c_n \nabla_{\mathbf{k}} \cdot \nabla_{\mathbf{x}_c} \mathcal{E}_n 
+ c_n \left( \mathcal{E}_n - \mathbf{x}_c \cdot \nabla_{\mathbf{x}_c} \mathcal{E}_n \right) \\ \nonumber
&+& c_m \biggl\{ \mathcal{E}^M_{mn} +  \frac{i}{2} \nabla_{\mathbf{x}_c} \left( \mathcal{E}_m + \mathcal{E}_n \right) \cdot \mathbf{\mathcal{A}}_{mn}  \biggl\}.
\end{myeqnarray}

Finally, by multiplying both sides of (\ref{eq4.1}) by $\sqrt{\rho}$ and integrating over $\mathbf{k}$, one finds

\begin{myeqnarray}\label{eq4.3}
i \hbar \dot z_n( \mathbf{k}_c ) &=& \left( \mathcal{E}_n - \mathbf{x}_c \cdot \nabla_{\mathbf{x}_c} \mathcal{E}_n \right) z_n 
+ \frac{i}{2} \left( \nabla_{\mathbf{x}_c} \mathcal{E}_n - \sum_{m \in D} | z_m |^2 \nabla_{\mathbf{x}_c} \mathcal{E}_m \right) \nabla_{\mathbf{k}_c} z_n \\ \nonumber
&+& \sum_{m \in I_D} \biggl\{ \mathcal{E}^M_{mn} +  \frac{i}{2} \nabla_{\mathbf{x}_c} \left( \mathcal{E}_m + \mathcal{E}_n \right) \cdot \mathbf{\mathcal{A}}_{mn} 
- i \hbar \langle u_n^0 | \dot u_m^0 \rangle \biggl\} z_m.
\end{myeqnarray}

These are the coupled dynamics for the degrees of freedom $z_n$ in the Bloch wave packet. 
The argument for each quantity on the right hand side is simply $ \mathbf{k}_c $.

\section{Near Degeneracy}

The above results require the subspace $D$ to be degenerate. Next we consider the near-degenerate case, where states in given Block band eigenstates with different wave vectors have close in energy eigenvalues. Schematically this situation can be given in Fig. S1, where the degenerate subspace $D$  and the near degenerate subspace $\tilde{D}$  are shown. The eigenstates of the unperturbed Hamiltonian $H^0$ are $\psi_n^0 \in D$ and $\psi_m^0 \in \tilde{D}$. Since $H^0$ is nearly degenerate in $\tilde{D}$, then $|E_n^0 - E_m^0| < \langle \psi_m^0 | H^1 | \psi_n^0 \rangle$ meaning that $|E_n^0 - E_m^0|$ is small with respect to $H^1$.

\begin{figure}
\includegraphics[scale=0.4]{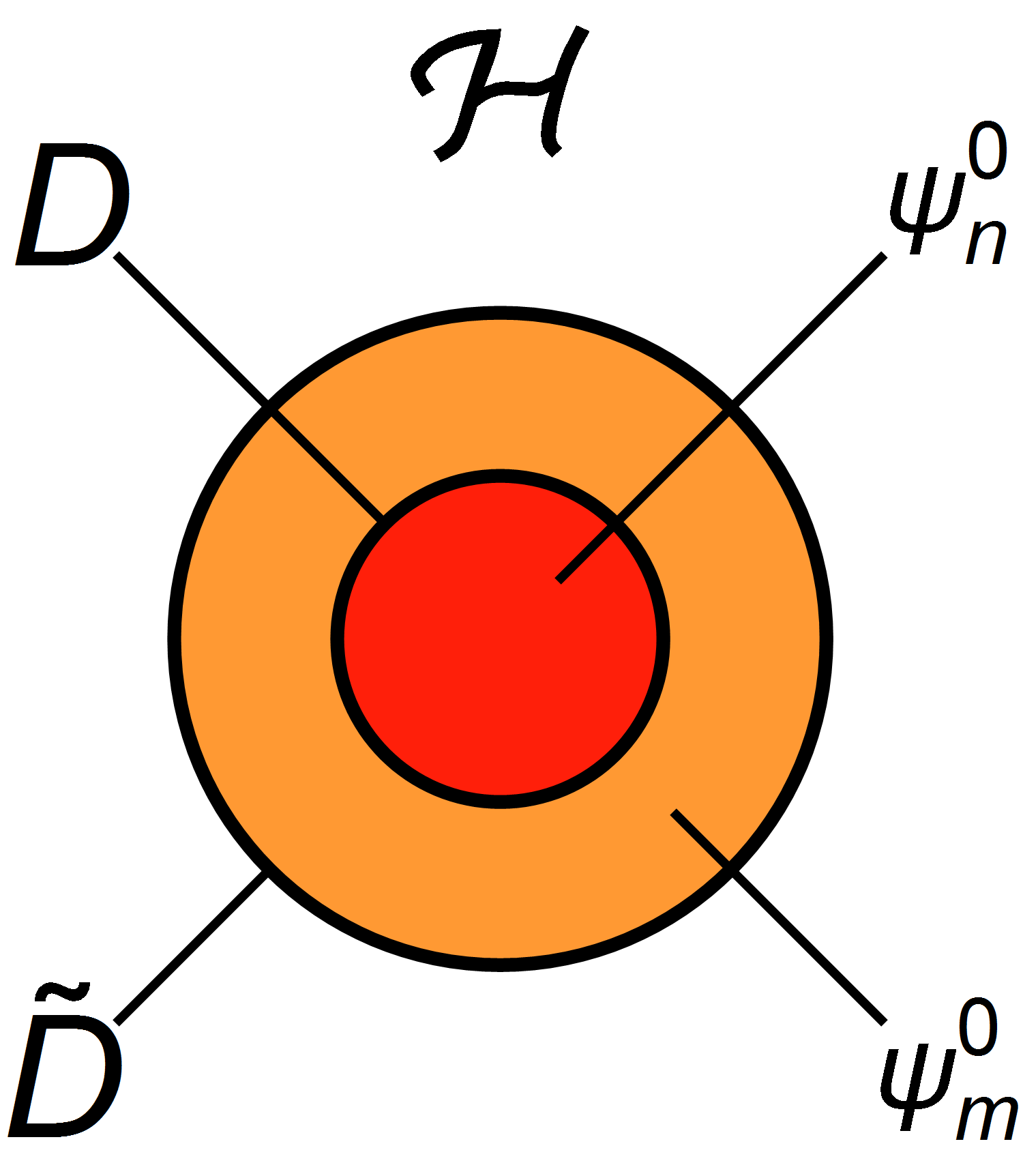}
\caption{Schematic representation of the degenerate subspace $D$ (red) with states $\psi_m^0$ and the larger space $\tilde{D}$ (orange) that includes states $\psi_n^0$ that are nearly degenerate to the states $\psi_m^0 \in D$.  The full Hilbert space is denoted $\mathcal{H}$.}
\label{fig}
\end{figure}

The full Hamiltonian is represented as $H = H^0 + H^1 = \bar{H}^0 + \Delta \bar{H} + H^1$. Here $\bar{H}^0$ has a degenerate spectrum, $\bar{H}^0 \psi_m^0 =\bar{E}^0 \psi_m^0$ for $\psi_m^0 \in \tilde{D}$, where $\bar{E}^0$ is an average of the energies for $\tilde{D}$. 
Also, $\bar{H}^0 \psi_m^0 = H^0 \psi_m^0 = E_m^0 \psi_m^0$ for $\psi_m^0 \notin \tilde{D}$. 
We then have, $[\bar{H}^0 , H^0] = 0$, and these operators are diagonal with matrix elements
$ \langle \psi_m^0 | \bar{H}^0 | \psi_n^0 \rangle = \bar{E}^0 \delta_{mn}$ for $\psi_m^0 , \psi_n^0 \in D$ and $ \langle \psi_m^0 | \bar{H}^0 | \psi_n^0 \rangle = E_n^0 \delta_{mn}$ otherwise. 
The diagonal Hermitian operator $\Delta \bar{H} = H^0 - \bar{H}^0$, with matrix elements $ \langle \psi_m^0 | \Delta \bar{H} | \psi_n^0 \rangle = \left( E_n^0 - \bar{E}^0 \right) \delta_{mn}$. 
Essentially, the full Hamiltonian $H$ contains a degenerate in $\tilde{D}$ unperturbed $\bar{H}^0$ and $\Delta \bar{H} + H^1$ are treated as a perturbation. 
However, for the degenerate $\bar{H}^0$, we can make use of the projected EOM from (\ref{eq1.24}) giving
\begin{myeqnarray}\label{eq2.1}
\dot{\langle O \rangle} = \frac{i}{\hbar} \left[ \mathcal{P}_D (H - \bar{H}^0) , \mathcal{P}_D (O) \right] + \langle \frac{d}{dt}  \mathcal{P}_D (O) \rangle. 
\end{myeqnarray}

However, $\mathcal{P}_D (\bar{H}^0) $ is proportional to the identity matrix in $D$ and hence commutes with any other operator in $D$, one obtains
\begin{myeqnarray}\label{eq2.2}
\dot{\langle O \rangle} = \frac{i}{\hbar} \left[ \mathcal{P}_D (H) , \mathcal{P}_D (O) \right] + \langle \frac{d}{dt}  \mathcal{P}_D (O) \rangle. 
\end{myeqnarray}
So, even in cases of near degeneracy, the projected EOM hold.

\section{Elements in the EOM Calculations}

The projected EOM in (5) and (6) of the main text within first order of the perturbation $H^1$ can now be used to obtain the time dynamics of the position $\mathbf{x}$ and the gauge invariant wave vector $ \mathbf{k} = \mathbf{q} - \frac{e}{\hbar}\mathbf{A}_c  $ found in  (7) of the main text. 

The wave packet subspace $D$ is given by the Bloch states $\psi_n (\mathbf{k})$ with $n\in I_D$, which is the natural basis for the calculations that follow. In this representation, the elements of position $\mathbf{x}$ and gauge invariant wave vector $\mathbf{k}$, needed for $\mathcal{P}_D(\mathbf x)$ and  $\mathcal{P}_D(\mathbf k)$, are found as
\begin{mysubeqnarray}\label{eq3.1}
\begin{align}
\langle \psi_{n_1} (\mathbf{k}) | \mathbf{x} | \psi_{n_2} (\mathbf{k}') \rangle &= \delta_{n_1n_2} \left( i\nabla_\mathbf{k} + \mathbf{\mathcal{A}}_{n_1n_2} \left( \mathbf{k} \right) \right) \delta \left( \mathbf{k}-\mathbf{k}' \right),   \\
\langle \psi_{n_1} (\mathbf{k}) | \mathbf{k} | \psi_{n_2} (\mathbf{k}') \rangle &= \delta_{n_1n_2} \mathbf{k} \delta \left( \mathbf{k}-\mathbf{k}' \right),
\end{align}
\end{mysubeqnarray}
where $\mathbf{\mathcal{A}}_{n_1n_2} (\mathbf{k}) =  i \langle u_{n_1} (\mathbf{k}) |   \nabla | u_{n_2} (\mathbf{k}) \rangle$. 
One also needs to consider $\mathcal{P}_D(H^1)$ where $H^1 = \frac{1}{2} \left( \frac{\partial H_c}{\partial \mathbf{x}_c} \cdot (\mathbf{x} - \mathbf{x}_c) + (\mathbf{x} - \mathbf{x}_c) \cdot \frac{\partial H_c}{\partial \mathbf{x}_c} \right)$. 
We find that 
\begin{myeqnarray}\label{eq3.2}
\langle \psi_{n_1} (\mathbf{k}) \left| \frac{\partial H_c}{\partial \mathbf{x}_c} \right| \psi_{n_2} (\mathbf{k}') \rangle = \delta \left( \mathbf{k}-\mathbf{k}' \right) \left[ \delta_{n_1n_2} \nabla_{\mathbf{x}_c} \mathcal{E}_{n_1} \left( \mathbf{k} \right) 
	 - \left( \mathcal{E}_{n_1} \left( \mathbf{k} \right) - \mathcal{E}_{n_2} \left( \mathbf{k}' \right) \right)\langle u_{n_1} \left( \mathbf{k} \right) | \nabla_{\mathbf{x}_c} u_{n_2} \left( \mathbf{k}' \right) \rangle \right].
\end{myeqnarray}

With these results, we can then obtain
\begin{myeqnarray}\label{eq3.3}
\langle \psi_{n_1} (\mathbf{k}) \left| \frac{\partial H_c}{\partial \mathbf{x}_c} \cdot \mathbf{x} \right| \psi_{n_2} (\mathbf{k}') \rangle &=& \nabla_{\mathbf{x}_c} \mathcal{E}_{n_1} \cdot \langle \psi_{n_1} (\mathbf{k}) | \mathbf{x} | \psi_{n_2} (\mathbf{k}') \rangle
	+ i \nabla_{\mathbf{k}'} \cdot \left[ \left( \mathcal{E}_{n_1} \left( \mathbf{k} \right) - \mathcal{E}_{n_2} \left( \mathbf{k}' \right) \right) \langle u_{n_1} \left( \mathbf{k} \right) | \nabla_{\mathbf{x}_c} u_{n_2} \left( \mathbf{k}' \right) \rangle \right] \nonumber \\
	&+& i \langle \nabla_{\mathbf{x}_c} u_{n_1} \left( \mathbf{k} \right) \left| \cdot \left( \mathcal{E}_{n_1} \left( \mathbf{k} \right) - H_c \left( \mathbf{k} \right) \right) \right| \nabla_{\mathbf{k}'}  u_{n_2} \left( \mathbf{k}' \right) \rangle  \delta \left( \mathbf{k}-\mathbf{k}' \right).
\end{myeqnarray}

Finally, the elements of the perturbation $H^1$ are calculated as
\begin{myeqnarray}\label{eq3.4}
\langle \psi_{n_1} (\mathbf{k}) \left| H^1 \right| \psi_{n_2} (\mathbf{k}') \rangle &=& \frac{1}{2} \nabla_{\mathbf{x}_c} \left( \mathcal{E}_{n_1} \left( \mathbf{k} \right) + \mathcal{E}_{n_2} \left( \mathbf{k}' \right) \right) \cdot \langle \psi_{n_1} (\mathbf{k}) | \mathbf{x} | \psi_{n_2} (\mathbf{k}') \rangle\nonumber \\
	&-& \frac{i}{2} \left( \nabla_{\mathbf{k}} - \nabla_{\mathbf{k}'} \right) \cdot \left[ \left( \mathcal{E}_{n_1} \left( \mathbf{k} \right) - \mathcal{E}_{n_2} \left( \mathbf{k}' \right) \right) \langle u_{n_1} \left( \mathbf{k} \right) | \nabla_{\mathbf{x}_c} u_{n_2} \left( \mathbf{k}' \right) \rangle \delta \left( \mathbf{k}-\mathbf{k}' \right) \right] \nonumber \\
	&+& \mathcal{E}^M_{n_1n_2} \delta \left( \mathbf{k}-\mathbf{k}' \right) - \mathbf{x}_c \cdot \langle \psi_{n_1} (\mathbf{k}) \left| \frac{\partial H_c}{\partial \mathbf{x}_c} \right| \psi_{n_2} (\mathbf{k}') \rangle.
\end{myeqnarray}

Here the interband magnetization contribution to (\ref{eq3.4}) is 
\begin{myeqnarray}\label{eq3.5}
\mathcal{E}^M_{n_1n_2} &=&  \frac{i}{2} \biggl\{ \langle \nabla_{\mathbf{x}_c} u_{n_1} \left( \mathbf{k} \right) \left| \cdot \left( \mathcal{E}_{n_1} \left( \mathbf{k} \right) - H_c \left( \mathbf{k} \right) \right) \right| \nabla_{\mathbf{k}'}  u_{n_2} \left( \mathbf{k}' \right) \rangle \nonumber \\
&-&\langle \nabla_{\mathbf{k}} u_{n_1} \left( \mathbf{k} \right) \left| \cdot \left( \mathcal{E}_{n_2} \left( \mathbf{k}' \right) - H_c \left( \mathbf{k} \right) \right) \right| \nabla_{\mathbf{x}_c}  u_{n_2} \left( \mathbf{k}' \right) \rangle \biggl\}.
\end{myeqnarray}

At this point, let us note that if a function $f(\mathbf{k})$ depends on $\mathbf{k}$, then $\nabla_{\mathbf{x}_c}f = -\frac{e}{\hbar} \sum_j \nabla_{\mathbf{x}_c} (\mathbf{A}_c)_j \frac{\partial}{\partial \mathbf{k}_j} f $ and $\frac{d}{dt}f = -\frac{e}{\hbar} \dot{\mathbf{A}}_c \cdot \nabla_{\mathbf{k}} f$. 
This observation applies to quantities like $u_{n_2} (\mathbf{k})$ and $\mathcal{E}_{n_2} (\mathbf{k})$, for example.
Therefore, (\ref{eq3.4}) becomes
\begin{myeqnarray}\label{eq3.5}
\mathcal{E}^M_{n_1n_2} &=&  -\frac{ie}{2\hbar} \sum_j \nabla_{\mathbf{x}_c} (\mathbf{A}_c)_j \biggl\{ \langle \nabla_{\mathbf{k}_j} u_{n_1} \left( \mathbf{k} \right) \left| \cdot \left( \mathcal{E}_{n_1} \left( \mathbf{k} \right) - H_c \left( \mathbf{k} \right) \right) \right| \nabla_{\mathbf{k}'}  u_{n_2} \left( \mathbf{k}' \right) \rangle \nonumber \\
&-&\langle \nabla_{\mathbf{k}} u_{n_1} \left( \mathbf{k} \right) \left| \cdot \left( \mathcal{E}_{n_2} \left( \mathbf{k}' \right) - H_c \left( \mathbf{k} \right) \right) \right| \nabla_{\mathbf{k}'_j}  u_{n_2} \left( \mathbf{k}' \right) \rangle \biggl\}.
\end{myeqnarray}

In the circular gauge $\mathbf{A} = \frac{1}{2} \left( \mathbf{B} \times \mathbf{x} \right)$, when $\mathbf{k}=\mathbf{k}'$, the above reduces to
\begin{myeqnarray}\label{eq3.6}
\mathcal{E}^M_{n_1n_2} = \mathbf{B} \cdot \mathbf{M}_{n_1n_2} (\mathbf{k}) =  - \frac{ie}{4\hbar} \mathbf{B} \cdot \langle  \nabla u_{n_1} (\mathbf{k})| 
\left( \mathcal{E}_{n_1} + \mathcal{E}_{n_2} - 2 H_c(\mathbf {k}) \right) \times |  \nabla  u_{n_2}(\mathbf{k}) \rangle,
\end{myeqnarray}
where we have defined the interband magnetization $\mathbf{M}_{n_1n_2} (\mathbf{k}) =  - \frac{ie}{4\hbar}\langle  \nabla u_{n_1} (\mathbf{k})| 
\left( \mathcal{E}_{n_1} + \mathcal{E}_{n_2} - 2 H_c(\mathbf {k}) \right) \times |  \nabla  u_{n_2}(\mathbf{k}) \rangle $, similar to what is found in the main text after (7). The other interband quantities in (7) of the main text are obtained by considering the following commutator, which can be calculated using (\ref{eq3.1}),

\begin{myeqnarray}\label{eq3.7}
\left[ \mathcal{P}_D (\mathbf{x}_i), \mathcal{P}_D (\mathbf{x}_j) \right] = i \sum_{n_1,n_2 \in D} \int d^3\mathbf{k} \left\{ (\Omega_{n_1n_2})_{ij} (\mathbf{k}) -i \sum_{n_3 \in I_D} (Q_{n_1n_3n_2})_{ij} (\mathbf{k}) \right\} \left| \psi_{n_1} (\mathbf{k}) \rangle \langle \psi_{n_2} (\mathbf{k}) \right|, 
\end{myeqnarray}

where

\begin{mysubeqnarray}\label{eq3.8}
\begin{align}
(\Omega_{n_1n_2})_{ij} (\mathbf{k}) &=  \frac{\partial}{\partial k_j } (\mathbf{\mathcal{A}}_{n_1n_2})_i (\mathbf{k}) -  \frac{\partial}{\partial k_i } (\mathbf{\mathcal{A}}_{n_1n_2})_j (\mathbf{k}) = i \left( \frac{\partial}{\partial k_j } \langle u_{n_1} (\mathbf{k}) | \frac{\partial}{\partial k_i } u_{n_2} (\mathbf{k}) \rangle  - \frac{\partial}{\partial k_ i} \langle u_{n_1} (\mathbf{k}) | \frac{\partial}{\partial k_j } u_{n_2} (\mathbf{k}) \rangle \right),  \\
(Q_{mkn})_{ij} (\mathbf{k}) &= (\mathbf{\mathcal{A}}_{n_1n_3})_i (\mathbf{k}) \left(\mathbf{\mathcal{A}}_{n_3n_2}\right)_j (\mathbf{k}) - (\mathbf{\mathcal{A}}_{n_1n_3})_j (\mathbf{k}) (\mathbf{\mathcal{A}}_{n_3n_2})_i (\mathbf{k}) \nonumber  \\
&= - \left( \langle u_{n_1} (\mathbf{k}) | \frac{\partial}{\partial k_ i} u_k (\mathbf{k}) \rangle  \langle u_k (\mathbf{k}) | \frac{\partial}{\partial k_j } u_{n_2} (\mathbf{k}) \rangle  
- \langle u_{n_1} (\mathbf{k}) | \frac{\partial}{\partial k_j } u_k (\mathbf{k}) \rangle  \langle u_k (\mathbf{k}) | \frac{\partial}{\partial k_ i} u_{n_2} (\mathbf{k}) \rangle   \right).
\end{align}
\end{mysubeqnarray}
These can be recast into a more familiar form found after (7) of the main text by defining their vector components as $(\mathbf{\Omega}_{n_1n_2})_i (\mathbf{k}) = \frac{1}{2}  \epsilon_{ijk} (\Omega_{n_1n_2})_{jk} (\mathbf{k}) $ and 
$(\mathbf{Q}_{n_1n_3n_2})_i (\mathbf{k}) = \frac{1}{2}  \epsilon_{ijl} (Q_{n_1n_3n_2})_{jl} (\mathbf{k}) $. 
In vector form, these become 

\begin{myeqnarray}\label{eq3.9}
\mathbf{\Omega}_{n_1n_2} (\mathbf{k}) &=& \nabla \times \mathbf{\mathcal{A}}_{n_1n_2} (\mathbf{k}), \nonumber \\
 \mathbf{Q}_{n_1n_3n_2} (\mathbf{k}) &=& \mathbf{\mathcal{A}}_{n_1n_3} (\mathbf{k}) \times \mathbf{\mathcal{A}}_{n_3n_2} (\mathbf{k}).
\end{myeqnarray}

The last terms in (5) and (6) of the main text involve the time derivatives of the projected operators. 
Using the above results, it is found that 
\begin{myeqnarray}\label{eq3.10}
\frac{d}{dt}\mathcal{P}_D (\mathbf{x})&=& i\frac{e}{\hbar} \sum_j \left(\dot{\mathbf{A}}_c\right)_j \left[ \mathcal{P}_D (\mathbf{x}_j) , \mathcal{P}_D (\mathbf{x}) \right] , \nonumber \\
\frac{d}{dt}\mathcal{P}_D (\mathbf{q})&=& 0.
\end{myeqnarray}

The results in (\ref{eq3.4}) and (\ref{eq3.10}) then give the time dynamics in (5) and (6) of the main text for the position $\mathbf{x}$ and gauge invariant wave vector $\mathbf{k}$ upon taking expectation values using a wave packet distribution $\rho (\mathbf{k}) = \delta (\mathbf{k} - \mathbf{k}_c)$.

We finally alert the reader that the interband properties in the main text were written as 
$\bar{\mathbf{\mathcal A}}_{n_1n_2}, \bar{\mathbf{\Omega}}_{n_1n_2}, \bar{\mathbf{M}}_{n_1n_2}, \bar{\mathbf{Q}}_{n_1n_2n_3}$ to denote that these were evaluated at $\mathbf {k}_c$, while in this document these interband properties are functions of the general $\mathbf {k}$.